\documentclass[twocolumn]{aastex62}

\renewcommand{\thefootnote}{\ifcase\value{footnote}
  \or *
  \or **
  \or $\dagger$
  \else \arabic{footnote}\fi}

\newcommand{\galname}{ESO~601--G036}
\newcommand{\ha}{H$\alpha$}
\newcommand{\hb}{H$\beta$}
\newcommand{\oiii}{[O~{\sc iii}]}

\newcommand{\nii}{[N~{\sc ii}]}
\graphicspath{{./}{figures/}}

\received{month date, 2018}
\revised{month date, 2018}
\accepted{\today}
\submitjournal{ApJ}

\shorttitle{The host galaxy of FRB~171020}
\shortauthors{Mahony et al.}

\begin{document}

\title{A search for the host galaxy of FRB~171020}

\correspondingauthor{Elizabeth Mahony}
\email{elizabeth.mahony@csiro.au}

\author[0000-0002-5053-2828]{Elizabeth K. Mahony}
\affil{Australia Telescope National Facility, CSIRO Astronomy and Space Science, PO Box 76, Epping, NSW 1710, Australia }

\author{Ron D. Ekers}
\affil{Australia Telescope National Facility, CSIRO Astronomy and Space Science, PO Box 76, Epping, NSW 1710, Australia }

\author[0000-0001-6763-8234]{Jean-Pierre Macquart}
\affil{International Centre for Radio Astronomy Research, Curtin University, Bentley WA 6102, Australia }

\author[0000-0002-1136-2555]{Elaine M. Sadler}
\affil{Australia Telescope National Facility, CSIRO Astronomy and Space Science, PO Box 76, Epping, NSW 1710, Australia }
\affiliation{Sydney Institute for Astronomy, School of Physics A28, The University of Sydney, NSW 2006, Australia}

\author[0000-0003-2149-0363]{Keith W. Bannister}
\affil{Australia Telescope National Facility, CSIRO Astronomy and Space Science, PO Box 76, Epping, NSW 1710, Australia }

\author{Shivani Bhandari}
\affil{Australia Telescope National Facility, CSIRO Astronomy and Space Science, PO Box 76, Epping, NSW 1710, Australia }

\author{Chris Flynn}
\affiliation{Centre for Astrophysics and Supercomputing, Swinburne University of Technology, John St, Hawthorn, VIC 3122, Australia}

\author[0000-0003-4351-993X]{B\"arbel S. Koribalski}
\affil{Australia Telescope National Facility, CSIRO Astronomy and Space Science, PO Box 76, Epping, NSW 1710, Australia }

\author{J. Xavier Prochaska}
\affil{University of California, Santa Cruz, 1156 High St., Santa Cruz, CA 95064, USA}
\affiliation{
Kavli Institute for the Physics and Mathematics of the Universe (Kavli IPMU),
5-1-5 Kashiwanoha, Kashiwa, 277-8583, Japan}

\author[0000-0003-4501-8100]{Stuart D. Ryder}
\affiliation{Department of Physics and Astronomy, Macquarie University, NSW 2109, Australia }

\author[0000-0002-7285-6348]{Ryan M. Shannon}
\affiliation{Centre for Astrophysics and Supercomputing, Swinburne University of Technology, John St, Hawthorn, VIC 3122, Australia}

\author[0000-0002-1883-4252]{Nicolas Tejos}
\affiliation{Instituto de F\'isica, Pontificia Universidad Cat\'olica de Valpara\'iso,
Casilla 4059, Valpara\'iso, Chile}

\author[0000-0003-1160-2077]{Matthew T. Whiting}
\affil{Australia Telescope National Facility, CSIRO Astronomy and Space Science, PO Box 76, Epping, NSW 1710, Australia }

\author{O.~I. Wong}
\affil{International Centre for Radio Astronomy Research-M468, The University of Western Australia, 35 Stirling Hwy, Crawley, WA 6009, Australia}



\begin{abstract}

We report on a search for the host galaxy of FRB~171020, the fast radio burst with the smallest recorded dispersion measure (DM $=114$\,pc\,cm$^{-3}$) of our on-ongoing ASKAP survey. The low DM confines the burst location within a sufficiently small volume to rigorously constrain the identity of the host galaxy. We identify 16 candidate galaxies in the search volume and single out \galname, a Sc galaxy at redshift $z=0.00867$, as the most likely host galaxy. UV and optical imaging and spectroscopy reveal this galaxy has a star-formation rate of approximately 0.1\,M$_\odot$\,yr$^{-1}$ and oxygen abundance $12 + \log({\rm O/H}) = 8.3 \pm 0.2$, properties remarkably consistent with the galaxy hosting the repeating FRB~121102. However, in contrast to FRB~121102, follow-up radio observations of \galname\ show no compact radio emission above a 5$\sigma$ limit of $L_{2.1\,{\rm GHz}}=3.6\times 10^{19}$ W\,Hz$^{-1}$. Using radio continuum observations of the field, combined with archival optical imaging data, we find no analog to the persistent radio source associated with FRB~121102 within the localization region of FRB~171020 out to $z=0.06$. These results suggest that FRBs are not necessarily associated with a luminous and compact radio continuum source. 

\end{abstract}

\keywords{galaxies: spiral --- galaxies: individual (\galname) --- radio continuum: galaxies}


\section{Introduction} \label{sec:intro}

The progenitors and local environments responsible for the bright \citep[$F_\nu \sim 0.55$\,--\,$380$~Jy~ms; ][]{Petroff2016}, extragalactic millisecond-duration pulses known as Fast Radio Bursts (FRBs) are open questions.  Only the single known repeating FRB \citep[121102; ][]{Spitler2016} has been localized sufficiently accurately to unambiguously identify its host galaxy \citep{Chatterjee2017,Tendulkar2017} and permit examination of the burst environment.   
Its host galaxy, a low-metallicity dwarf galaxy at redshift $z=0.193$, harbors a `persistent' compact radio source \citep{Chatterjee2017} co-located within 40~pc of the FRB \citep{marcote2017}.  The radio source, and therefore the FRB, likely resides in a bright star-forming region in the outskirts of the galaxy \citep{bassa2017}. The persistent radio source has inspired suggestions that detections of bright radio emission may help to identify hosts 
of other FRBs \citep{eftekhari2018}.

However, the unusual properties of FRB\,121102 cast doubt on 
applying these findings to other FRBs.  So far, no other FRBs are observed to repeat, 
despite extensive follow-up campaigns \citep[e.g.][]{Bhandari2018}. Radio diagnostics of the repeater's host galaxy and circum-burst medium reveal that FRB\,121102 resides in a highly magnetized medium \citep{Michillietal2018} whose rotation measure exceeds other FRB measurements by 3--4 orders of magnitude \citep{Masuietal2015,Petroffetal2017,Calebetal2018}. 

Here we examine FRB~171020 which has the lowest DM measured
to date \citep[114\,pc\,cm$^{-3}$;][]{shannon2018}. No repeat bursts above S/N of 9 were found in 32.7d of observations. Despite the large localization region of this FRB, 
$50 \times 34$ arcmin  at a position angle of 29.6$^\circ$ (95\% containment), its low DM demands a sufficiently close
proximity to attempt identification of its host galaxy. 

In Section~2 we report the properties of FRB~171020 and estimate the maximum redshift, followed by a search for its host galaxy in existing catalogs (Section~3). We discuss follow-up optical and radio observations of our best candidate host galaxy \galname\ in Section~4 and compare the properties of this galaxy with the host galaxy of FRB~121102 in Section~5 before concluding in Section~6. The cosmological parameters assumed in this paper are from the Planck 2015 results \citep{planck2016}. 

\section{The maximum redshift of FRB~171020} \label{sec:dm}

We use the dispersion measure of FRB~171020 to estimate an upper limit to its distance. 
We assume that the total DM is given by:
$$\mathrm{DM}=\mathrm{DM_{MW(disk)}} + \mathrm{DM_{MW(halo)}} + \mathrm{DM_{IGM}} + \mathrm{DM_{Host}}$$ 

\noindent where $\mathrm{DM_{MW(disk)}}$ and $\mathrm{DM_{MW(halo)}}$ are the contributions, respectively, of the Milky Way disk ISM and halo, $\mathrm{DM_{IGM}}$ is the contribution of the intergalactic medium along the line of sight, and DM$_{\rm Host}$ is the contribution from both the host galaxy itself and its circum-burst environment. Given the DM of FRB~171020 is so low the contribution from the Milky Way represents a significant fraction of the total. As such, the assumptions used to derive these quantities can lead to large fractional
uncertainty in the maximum redshift of the FRB. 

At the Galactic coordinates $(l, b) = (36.4,-53.6)$ of the burst, DM$_{\mathrm{MW(disk)}}$ is 38\,pc\,cm$^{-3}$ according to the NE2001 model \citep{ne2001}, or 26\,pc\,cm$^{-3}$ following the YMW16 model \citep{ymw16}. 

The Milky Way halo contribution is much more uncertain; 
we assume $\rm DM_{\mathrm{MW(halo)}} = 12\, pc\,cm^{-3}$ 
calculated by using the excess DM of pulsars detected on the near side 
of the Large Magellanic Cloud \citep{PSRCAT}. By selecting the closest LMC pulsars it is assumed that there is negligible DM contribution from the LMC itself and subtracting the Milky Way contribution leads to a halo contribution of $8< $DM$_{\mathrm{MW(halo)}} <13$ \,pc\,cm$^{-3}$ out to 50\,kpc. 
Extragalactic FRBs, however, travel farther implying
a higher 
$\rm DM_{\mathrm{MW(halo)}}$ 
that depends on its electron density distribution. 
Physically plausible models predict 
$\rm DM_{\mathrm{MW(halo)}}$ 
with an additional 2--21\,pc\,cm$^{-3}$ 
at distances $>50$ kpc (e.g., 
\citealt{MillerBregman2013}, Prochaska \& Zheng, in prep.). This is consistent with the estimate of $\rm DM_{\mathrm{MW(halo)}} = 30$\,pc\,cm$^{-3}$ \citep{dolag2015}, but we use a lower estimate here to determine the maximum redshift. 

For $\rm DM_{\mathrm{Host}}$,
we adopt two different assumptions. 
One assumes $\rm DM_{\mathrm{Host}} = 0 \rm \,pc\,cm^{-3}$,
and the other assumes 
DM$_{\mathrm{Host}}=45$\,pc\,cm$^{-3}$ typical for a dwarf 
galaxy \citep{Xu2015}. Note that $\rm DM_{\mathrm{Host}}$ includes the halo, disk and circum-burst environment components of the host galaxy.

The two assumptions imply a range for DM$_{\rm IGM}$ 
of 0--76\,pc\,cm$^{-3}$ (Table \ref{DMtable}). 
Adopting the relation between redshift and DM$_{\mathrm{IGM}}$ of $z \approx $ DM$_{\mathrm{IGM}}/1000$ \citep{Ioka2003,Inoue2004} 
yields an upper limit on the redshift of $0.02 \lesssim z_{\rm max} \lesssim 0.08$. 
Given the low observed DM of FRB~171020, the contribution from the IGM could vary significantly depending on the number and properties of intervening galaxy halos \citep{McQuinn2014}. However, given that the maximum redshift of $z=0.08$ assumes conservative estimates of both the Milky Way halo and host galaxy contributions, any scatter in the DM-$z$ relation is compensated for by the large range in maximum redshifts obtained by using the different models in Table \ref{DMtable}.

\begin{table}
\caption{Estimates for the DM contributions (in pc\,cm$^{-3}$) from the Milky Way (MW) and FRB host galaxy, as described in \S\ref{sec:dm}\ of the text. Final row gives the estimated upper limit on the host galaxy redshift. \label{DMtable}}

\begin{tabular}{lrrrrrr}
&&& \\
& \multicolumn{2}{c}{\bf YMW16} & \multicolumn{2}{c}{\bf NE2001} \\
{\bf Model} & {\bf (a)} & {\bf (b)} & {\bf (c)} & {\bf (d)} \\
 \hline
 DM$_{\rm total, obs}$ & 114 & 114 & 114 & 114 \\
\hline
 DM$_{\rm MW(disk)}$ & 26 & 26 & 38 & 38  \\
 DM$_{\rm MW(halo)}$ & 12 & 12 & 12 & 12 \\
 DM$_{\rm Host}$     & 0 & 45 &  0 & 45 \\
 DM$_{\rm IGM}$      & 76 & 31 & 64 & 19 \\
\hline 
$z_{\rm est}$ & $0.08$ &  $0.03$  & $0.06$ & $0.02$ \\
\hline
\end{tabular}
\end{table}

\section{Candidate host galaxies of FRB~171020} \label{sec:opticalsearch}

The area of the localization region of FRB~171020 (0.38 deg$^2$) and its maximum distance of 350\,Mpc (using $z_{\rm max}=0.08$ from Model (a) in Table 1) together confine its host galaxy to a maximum comoving volume of 1620\,Mpc$^3$.  Taking the lower value of $z_{\rm max}=0.03$ from Model (b), 
the search volume is only 90\,Mpc$^3$. 
These volumes are small enough 
to search for an optical host galaxy counterpart to FRB~171020 in spite of 
its poor localization. 
We searched the NASA Extragalactic Database for cataloged galaxies within the FRB localization ellipse, yielding only two galaxies with a published redshift at $z<0.08$.  

\subsection{\bf \galname} 
\galname\ is a $\rm{B_J}=15.6$ Sc galaxy \citep{esouppsala} at a redshift of $z=0.00867$ measured from \mbox{H\,{\sc i}} emission detected in the \mbox{H\,{\sc i}} Parkes All Sky Survey (HIPASS; \citealt{hipass}). 
\cite{dacosta1998} list an optical radial velocity of 2539 km\,s$^{-1}$ ($z=0.0085$) giving a distance of 37\,Mpc using the
the \cite{Mould2000} model. 
The absolute magnitude is $M_{\rm R}=-17.9$, calculated from the SuperCOSMOS R-band after correcting for Galactic extinction. 

The probability that \galname\ is a chance association may be 
estimated from the surface density of nearby galaxies.
The Compact Binary Coalescence Galaxy (CBCG) catalog \citep{cbcg} 
list all star-forming galaxies to $z \sim 0.025$ 
including \galname. The 0.38\,deg$^2$ localization region for 
FRB~171020 gives a $\sim 40$\% chance of finding a CBCG galaxy within the $2\sigma$ ellipse decreasing to a 10\% probability if located in the $1\sigma$ localization area.

To expand this analysis out to higher redshifts we follow the method described in \citet{Eftekhari2017} to calculate the probability that \galname\ is associated with FRB~171020. The large localization region yields a probability of a chance coincidence to be 1, meaning that \galname\ is far from a statistically robust identification beyond $z \sim 0.025$. In the following sections we search for other possible counterparts within the search volume. 

\subsection{\bf 2MASX~J22150112$-$1925373} 
This is an elliptical galaxy at $z=0.0667$ \citep{6df} and absolute magnitude calculated from the SuperCOSMOS R-band magnitude of $M_{\rm R}=-21.5$. 
At this redshift, the average DM$_{\rm IGM}$ implies 
DM$_{\mathrm{host}}<10$\,pc\,cm$^{-3}$ which 
is implausible for this massive elliptical \citep{Xu2015, walker2018}.
This inconsistency and its location at the edge
of the 95\% confidence region make 2MASX~J22150112$-$1925373 an
unlikely host of FRB~171020.

\subsection{Candidates from the WISE$\times$SCOSPZ Catalog}

Since existing redshift catalogs are incomplete for low-luminosity galaxies, we expanded our search 
by using the WISE$\times$SuperCOSMOS Photometric Redshift Catalog (WISE$\times$SCOSPZ; \citealt{WISExSCOSPZ}). 
The catalog magnitude limit includes LMC-like galaxies 
(M$_{\rm R}=-18.5$) out to $z\sim0.08$, but is incomplete to 
dwarfs beyond $z\sim0.03$. 

We found sixteen objects with photometric redshifts $z_{\rm ph}<0.08$ within the localization region of FRB~171020, including \galname. The other fifteen objects all have photometric redshifts above $z=0.04$, and are listed in Table \ref{candidates_tab}.
We conducted follow-up observations of five of these candidates with 
the X-Shooter spectrograph \citep{Vernet2011} mounted on UT2 (Kueyen) of the
European Southern Observatory's Very Large Telescope  
on 2018~Aug~3~UT. 
Each source was observed at 2 nod positions for a total on-source integration time of
360\,s, through slit widths of $0\farcs9$ in the NIR
and VIS arms, and $1\farcs0$ in the UVB arm. 

A spectroscopic redshift was obtained for only 1 of these targets: WISEJ221621.59$-$191829.9 at $z=0.024$. While confirming this object is at $z<0.03$ makes it a more likely host galaxy candidate, it is also close to the edge of the $2\sigma$ error ellipse which decreases the probability it is the correct identification. No emission lines or significant stellar continuum were detected in either the UVB or VIS spectra of the other four sources, indicating that they are most likely beyond our 
search volume. 

\begin{table*}
\caption{List of all candidate host galaxies within the FRB171020 error ellipse. Where available, optical and IR magnitudes are taken from the WISE$\times$SCOSPZ catalog and have been extinction corrected. Otherwise magnitudes were obtained directly from the WISE and SuperCOSMOS databases (and not corrected for extinction). We report photometric redshifts from the WISE$\times$SCOSPZ to 2 s.f. (denoted by $\dagger$), but it is unlikely they are accurate to this level of significance. Source names marked by $\ddagger$ have been followed up spectroscopically with X-Shooter. A redshift was unable to be measured for six of these sources indicating that they are outside the search volume. The 2.1\,GHz flux densities are measured from the deeper ATCA observations carried out on 2018 June 28. If undetected, 5$\sigma$ limits measured at that position are listed. \label{candidates_tab}}
\begin{tabular}{llrccccclllc}
\hline
\# & Name & Prob.\footnote{The probability density (scaled to the peak pixel) at that position in the localization region calculated by the detection of the FRB in multiple beams. See \citet{Bannisteretal2017} for further details.} & $S_{\rm 2.1 GHz}$ & W1 & W2 & B & R & \multicolumn{1}{c}{$z$} & M$_{\rm R}$ & Notes \\
& & & (mJy) & (mag) & (mag) & (mag) & (mag) & & (mag) \\
\hline
\multicolumn{9}{l}{\emph{\bf NED matches out to $z=0.08$}} \\
1 & ESO 601$-$G036$\ddagger$ & 1 & 0.3 & 14.7 & 14.6 & 15.0 & 15.0 & 0.00867 & $-17.9$ &  \\
2 & 2MASX J22150112$-$1925373$\ddagger$ & 0.24 & 0.66 & 12.6 & 12.5 & 14.3 & 15.8 & 0.0667 & $-21.5$ & \\ 
\multicolumn{9}{l}{\emph{\bf WISE$\times$SCOSPZ matches out to $z=0.08$}} \\
1 & J221524.61$-$193504.8 & 1 & 0.3 & 14.7 & 14.6 & 15.0 & 15.0 & 0.00867 & $-17.9$ & ESO 601$-$G036 \\
3 & J221621.59$-$191829.9$\ddagger$ & 0.23 & $<0.08$ & 15.0 & 15.0 & 17.5 & 17.0 & 0.024 & $-18.2$ &  \\
4 & J221548.31$-$192225.0$\ddagger$ & 0.34 & $<0.07$ & 16.6 & 16.2 & 18.8 & 18.1 & 0.043$\dagger$ & $-18.4$ \\
5 & J221601.96$-$193251.4$\ddagger$ & 0.48 & $<0.09$ & 15.7 & 15.5 & 17.8 & 17.6 & 0.055$\dagger$ & $-19.4$ \\
6 & J221638.72$-$192651.0 & 0.54 & $<0.21$ & 14.1 & 14.0 & 17.1 & 16.5 & 0.055$\dagger$ & $-20.5$\\
7 & J221413.69$-$194032.1$\ddagger$ & 0.19 & $<0.24$ & 16.6 & 16.5 & 19.0 & 18.3 & 0.056$\dagger$ & $-18.8$ \\
8 & J221445.43$-$194502.2$\ddagger$ & 0.57 & $<0.09$ & 16.4 & 15.9 & 18.3 & 18.4 & 0.056$\dagger$ & $-18.7$ \\ 
9 & J221649.58$-$192707.1 & 0.26 & $<0.26$ & 16.5 & 16.1 & 18.8 & 18.2 & 0.057$\dagger$ & $-18.9$ \\
10 & J221437.97$-$192453.2 & 0.13 & $<0.27$ & 16.5 & 15.9 & 18.7 & 17.9 & 0.068$\dagger$ & $-19.6$ & \\
11 & J221611.23$-$191443.8 & 0.18 & $<0.27$& 16.0 & 15.8 & 18.5 & 17.8 & 0.066$\dagger$ & $-19.7$ & \\
12 & J221559.40$-$192629.3 & 0.53 & $<0.09$ & 14.1 & 14.1 & 17.3 & 16.2 & 0.070$\dagger$ & $-21.3$ & \\
13 & J221449.49$-$192207.4 & 0.04 & $<0.21$ & 16.0 & 15.6 & 18.2  & 18.1 & 0.071$\dagger$ & $-19.5$ \\ 
14 & J221528.16$-$193851.9 & 0.85 & $<0.08$ & 16.2 & 16.2 & 18.7  & 18.5 & 0.076$\dagger$ & $-19.3$ \\         
15 & J221503.26$-$192544.4 & 0.24 & $<0.21$ & 14.2 & 14.1 & 17.4  & 16.8 & 0.079$\dagger$ & $-21.1$ \\ 
16 & J221612.70$-$192222.1 & 0.52 & $<0.12$ & 15.6 & 15.5 & 18.4 & 17.5 & 0.076$\dagger$ & $-20.3$ \\    
17 & J221618.08$-$194206.2 & 0.09 & $<0.18$ & 14.3 & 14.0 & 17.3 & 16.9 & 0.079$\dagger$ & $-20.9$ \\   
\multicolumn{11}{l}{\emph{\bf Possible FRB~121102 analogs: $S_{\rm 2.1 GHz}>2.5$\,mJy with optical counterparts in SuperCOSMOS. }} \\
18 & J221430$-$195511 & 0.26 & 6.6 & 14.5 & 14.2 & 21.0 & 19.6 & -- & -- & LIRG WISE colors\\
19 & J221507$-$194713$\ddagger$ & 0.35 & 2.6 & 15.2 & 15.2 & 21.1 & 19.3 & $>0.1$ & -- & \\
20 & J221510$-$194835 & 0.35 & 2.8 & 15.2 & 15.0 & 19.6 & 18.6 & 0.17$\dagger$ & $-21.1$ &  \\
21 & J221525$-$194518 & 0.62 & 5.4 & 16.4 & 15.7 & 21.1 & 20.2 & -- & -- & Seyfert WISE colors  \\
22 & J221606$-$194032 & 0.26 & 4.0 & 15.5 & 14.1 & 20.1 & 19.2 & -- & & QSO WISE colors \\
23 & J221612$-$194915$\ddagger$ & 0.13 & 4.2 & 15.0 & 14.7 & 21.1 & 19.1 & $>0.1$ & -- & \\
24 & J221631$-$191942 & 0.30 & 3.2 & 14.2 & 13.9 & 20.1 & 18.1 & 0.35$\dagger$ & $-23.3$ & \\ 
\hline
\end{tabular}
\end{table*}

\section{Multi-wavelength properties of \galname}
Given the low redshift and location close to the center of the localization ellipse, \galname\ is the most likely host galaxy of FRB~171020. In low-resolution images of this galaxy (i.e. DSS) there is a clear stellar tail just south of the galaxy (separately cataloged by \cite{esouppsala} as ESO~601$-$G037). Bright UV emission of both objects, detected in GALEX images, indicates significant star formation which likely originated from tidal interactions or accretion of a dwarf galaxy companion. In the infrared, \galname\ is detected in the VISTA Hemisphere Survey with $Y=14.53, J=14.31, K{\rm s}=13.73$ mag \citep{vista}. We estimate a total star formation rate of $\sim0.13$ M$_\odot$\,yr$^{-1}$ from the {GALEX} data\footnote{This calculated SFR includes Galactic dust reddening and internal dust corrections as per \citet{Wong2016}}, and a total stellar mass of $\sim9\times10^8$ M$_\odot$ from the VISTA $K{\rm s}$-band magnitude.

The \mbox{H\,{\sc i}} Parkes All Sky Survey (HIPASS; \citealt{Barnes2001}) shows \mbox{H\,{\sc i}} emission in \galname, 
with an \mbox{H\,{\sc i}}-integrated flux density of 
$\approx 7\, \rm Jy\,km\,s^{-1}$, corresponding to an \mbox{H\,{\sc i}} mass of $2.3 \times 10^9$ M$_\odot$\footnote{These galaxy properties have been extracted from the HIPASS datacubes and differ slightly from those published by \citet{hipass}.}. For a rotational velocity of $\sim$80 km\,s$^{-1}$ and maximum radius of 11 kpc, we derive a dynamical mass of $\sim 1.6 \times10^{10}$ M$_\odot$.  

\galname\ is a member of a loose galaxy group, with four other gas-rich members detected in HIPASS at similar velocities. These other four galaxies lie outside the FRB~171020 2$\sigma$ error region. 

\subsection{Optical follow-up of \galname}

We observed \galname\ using the GMOS spectrograph \citep{Hook2004, Gimeno2016} on Gemini-South on 2018 July~11 UT. We used the B600 grating with a $0\farcs75$ wide slit oriented along its major axis for a total of $5$\,min exposure time with wavelength coverage of $3700-6850$\,\AA\ and dispersion of $\approx 1$\,\AA/pixel. 

The GMOS spectrum gives a redshift measurement of $z \approx 0.0082$, 
consistent with the \mbox{H\,{\sc i}} observations. 
We see strong \ha, [O\,{\sc iii}] emission lines, but little \nii\ emission indicating that this is a low-metallicity galaxy. 
The flux ratios of $\log ($\nii$\lambda6583$/\ha$)~\approx -1.09$ and $\log$\{(\oiii$\lambda5007$/\hb)/(\nii$\lambda6583$/\ha)\}~$\approx 1.41$,
imply an oxygen abundance $12 + \log({\rm O/H}) \approx 8.3 \pm 0.2$  
\citep{PettiniPagel2004}.  
This is consistent with the upper limit of $< 8.4$ found for the 
host galaxy of FRB~121102 \citep{Tendulkar2017}. 
The flux line ratios $\log ($\oiii$\lambda5007$/\hb$)\approx 0.32$
and $\log($\nii$\lambda6583$/\ha$)\approx -1.09$ place this galaxy in the 
star-forming region of the BPT diagram \citep{BPT1981}. 

In addition, we obtained narrow-band imaging of \galname\ with GMOS on Gemini-South on 2018 July~15. We observed for an exposure time of $3\times 180$\,s using the HaC filter ($6590-6650$\,\AA) ensuring full coverage of the \ha\ emission from this galaxy at redshift $z\sim 0.008$
(Figure \ref{atca_obs2}).


\subsection{Radio continuum follow-up of \galname}

\begin{figure*}
\centering{\includegraphics[width=\linewidth]{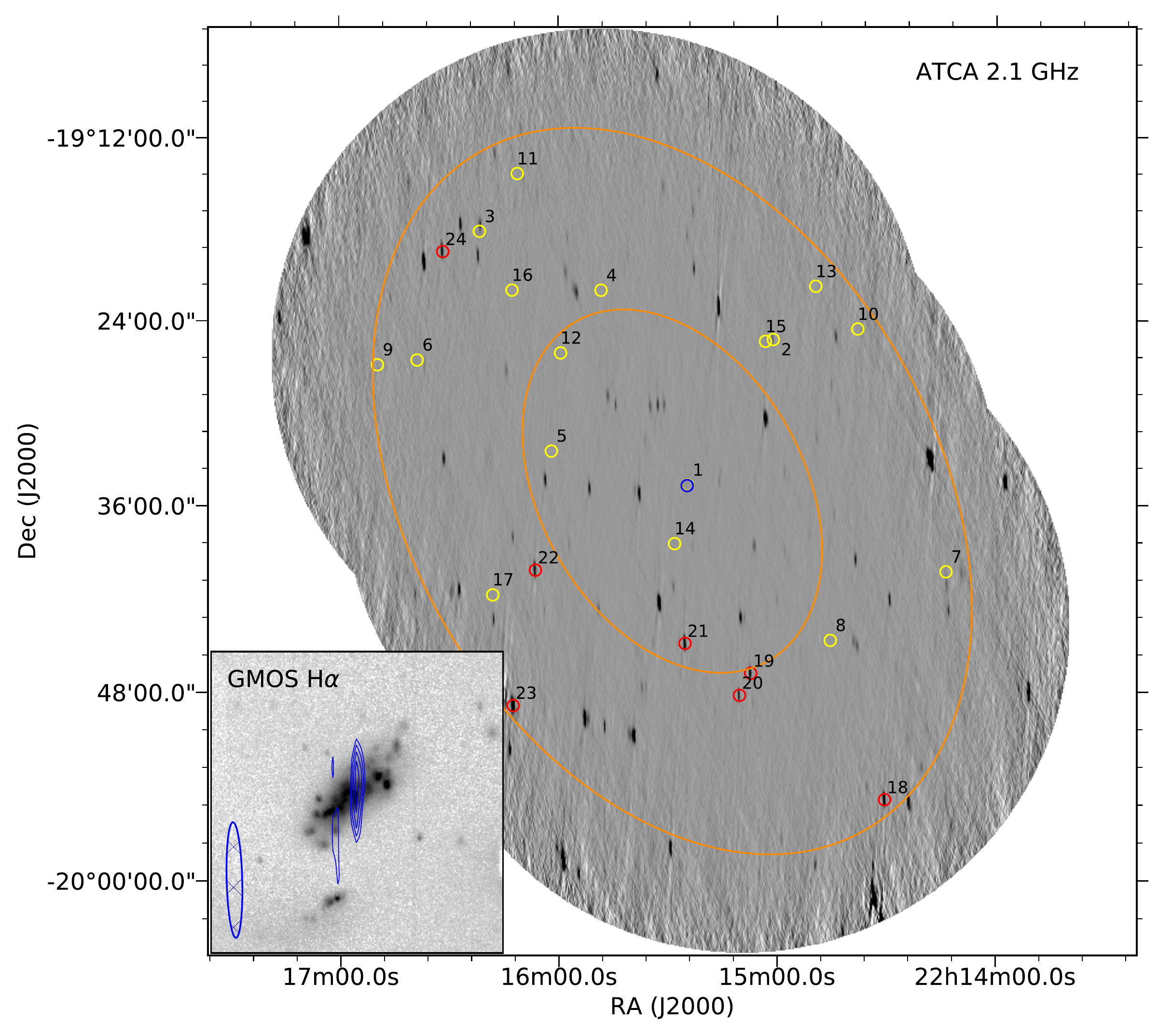}}
\caption{ATCA radio continuum image of the field of FRB~171020 at 2.1 GHz.  The orange ellipses mark the 1$\sigma$ and 2$\sigma$ confidence localization regions of FRB~171020. Yellow circles are optically-selected candidates from the WISE$\times$SCOSPZ catalog listed in Table \ref{candidates_tab}, red circles are radio-selected candidates (FRB~121102 analogs) discussed in Section~\ref{sec:radiosearch}. The blue circle highlights the position of ESO 601$-$G036. Inset: GMOS H$\alpha$ image of the galaxy \galname\ overlaid with a deep ATCA 2.1 GHz radio continuum emission shown in blue contours. The contour levels are 0.08, 0.09, 0.1, and 0.11 mJy\,beam$^{-1}$. The ATCA synthesized beam ($37\farcs1 \times 5\farcs0$) is shown in the bottom left corner. Candidate sources are labeled as in Table \ref{candidates_tab}.  \label{atca_obs2}}
\end{figure*}

\begin{figure}
\centering{\includegraphics[width=\linewidth]{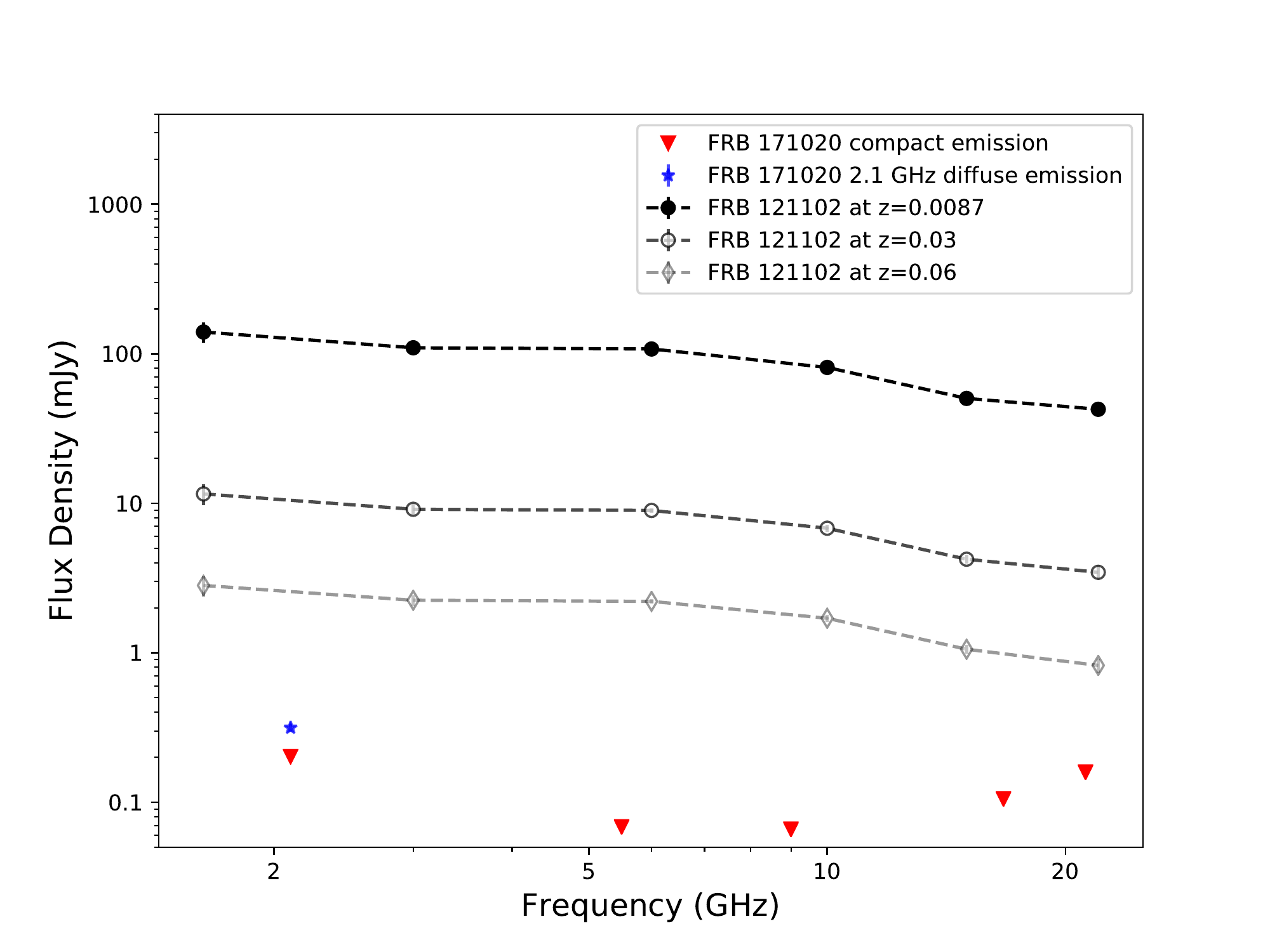}}
\caption{Flux density limits reached from ATCA observations of \galname. The red triangles denote 5$\times$rms values. Black points show the flux density of the persistent radio source detected in the repeating FRB if it were observed at $z=0.0087$ (the redshift of \galname), $z=0.03$ (open circles) and $z=0.06$ (open diamonds). The blue star shows the integrated flux density of the extended emission detected in deeper 2.1\,GHz observations. Error bars are plotted, but are generally smaller than the data points given the large range in flux density shown. \label{flux_limits}}
\end{figure}

To search for a `persistent' radio source, we carried out radio continuum observations using the Australia Telescope Compact Array (ATCA). The initial observations were carried out on 2018 May 27 UT across a wide frequency range to search for a compact radio source in ESO 601$-$G036 and to
compare its SED with the persistent source of FRB~121102. 

Using an extended ATCA array (6D configuration) and Briggs weighting with robust=0.5 resulted in synthesized beam sizes ranging from $17\farcs4 \times 4\farcs0$ at 2.1\,GHz to $1\farcs8 \times 0\farcs3$ at 21.2\,GHz. This gives rms noise levels of 40.5, 13.7, 13.2, 21.1 and 31.8 $\mu$Jy at frequencies of 2.1, 5.5, 9.0, 16.7 and 21.2\,GHz, respectively\footnote{The rms was calculated from the central 10\% of the primary-beam corrected image at each frequency except at 2.1\,GHz where the central 3\% was used to avoid nearby sources.}. No radio continuum emission associated with ESO 601$-$G036 was detected at any frequency. 

Figure \ref{flux_limits} shows the $5\sigma$ flux density limits reached at each frequency. 
If \galname\ hosts this FRB, 
there is no coincident radio source above a radio luminosity of $L_{\rm 2.1 GHz}=3.6\times 10^{19}$ W\,Hz$^{-1}$ ($5\sigma$), 
i.e. $600$ times fainter than that seen in the repeating FRB. 

Subsequent observations were carried out over the entire localization area at 1--3\,GHz  using a more compact array configuration (1.5D) on 2018 June 28 UT. The increased sensitivity (rms$=13.0\,\mu$Jy) and lower resolution of these observations revealed a faint continuum source at the center of ESO 601$-$G036 with a flux density of $S_{\rm 2.1 GHz}=240\,\mu$Jy (with a beam size of $37\farcs1 \times 5\farcs0$). Re-imaging using natural weighting gives a beam size of $83\farcs8 \times 12\farcs3$ and a flux density of $S_{\rm 2.1 GHz}=315\,\mu$Jy and indicates 
that the source is  resolved. Using the naturally-weighted flux density gives a radio luminosity of $L_{\rm 2.1 GHz}=4.3\times 10^{19}$ W\,Hz$^{-1}$. As this radio detection is resolved we discount it as being similar to the persistent radio source detected in FRB~121102, which is compact on mas-scales. 
Figure \ref{atca_obs2} shows the 2.1\,GHz data of the field. 
None of the candidates selected from the WISE$\times$SCOSPZ catalog 
have associated radio emission 
($5\sigma$ limits are listed in Table \ref{candidates_tab}). 

\section{Comparison of \galname\ with the host galaxy of FRB 121102}

Table \ref{tab_compare} compares key properties of \galname\ with those of the host galaxy of the repeating FRB~121102.
DM$_{\rm host}$ was estimated by assuming all excess DM is 
attributed to the host galaxy. 

\galname\ is about a magnitude more luminous than the host galaxy of FRB~121102 \citep{Tendulkar2017}, and the measured projected size of the galaxy ($9.2\times 3.7$\,kpc) is slightly larger. The current star-formation rate in \galname\ is two-three times lower than in the FRB~121102 host, but the two galaxies are qualitatively similar. 

The most striking difference is that \galname\ does not contain a luminous persistent radio source like that seen in the FRB~121102 host galaxy. If \galname\ is indeed the host galaxy of FRB~171020, this would imply that not all FRBs are associated with bright, compact and persistent radio emission. 

\begin{table}
\caption{Comparison of \galname\ and the host galaxy of the repeating FRB~121102. \label{tab_compare}}
\begin{tabular}{lrrr}
 \hline
 & \galname & FRB~121102   \\
 &          & host galaxy     \\
 \hline
Hubble type & Sc & Dwarf & \\
Redshift & 0.008672 & 0.19273  \\
D$_{\rm L}$ (Mpc) & 37 & 972   \\
 && \\
 M$_{\rm R}$ (mag) & $-$17.9 & $-$17.0  \\
SFR (M$_\odot$\,yr$^{-1}$) &  0.13   & 0.23--0.4  \\
Stellar Mass (M$_\odot$) & $9\times10^{8}$ & $1\times10^{8}$ \\
$\log ($\nii$\lambda6583$/\ha$)$ &$ -1.09$ & $\leq -1.34$\\
$12 + \log({\rm [O/H]})$ & $ 8.3 \pm 0.2$ & $< 8.4$\\
&&& \\
Radio continuum & $<3.6\times10^{19}$ & $2.3\times10^{22}$ \\
(W\,Hz$^{-1}$) &   \\
Host galaxy DM & $\sim64-76$ & $55-255$\\
(pc cm$^{-3}$)& & \\
\hline
&&& \\
\end{tabular}
     \raggedright {\bf Notes:} The values quoted for the FRB 121102 host galaxy are from \cite{Tendulkar2017,bassa2017}\\
\end{table}

\subsection{Searching for FRB~121102 analogs} \label{sec:radiosearch}

Since there is no evidence of a compact, persistent radio source associated with \galname, we consider whether there are other sources in the field that have similar properties to the host galaxy of FRB~121102, but may have been missed by the optical catalogs used in Section~\ref{sec:opticalsearch}. In the SuperCOSMOS passbands the host galaxy of FRB~121102 has optical magnitudes B $=26.2$ and R $=25.2$ meaning it would be detected above the SuperCOSMOS magnitude limits out to $z=0.06$. At this redshift, the persistent radio source  would be detected above $\sim 2.5$\,mJy at 2.1\,GHz and be brighter than $S_{\rm 2.1\,GHz}\sim 10$\,mJy if it was at $z<0.03$. 

There are 23 radio sources with $S_{\rm 2.1\,GHz}>2.5$\,mJy detected in the localization region of FRB~171020, but only seven of these are also detected in SuperCOSMOS. 
Of these seven radio sources, listed in Table \ref{candidates_tab}, two are cataloged in WISE$\times$SCOSPZ with a photometric redshift $z_{\rm ph}>0.1$, and three have WISE colors consistent with Luminous Infra-Red Galaxies (LIRGs) or QSOs indicating that these are likely background AGN.
The remaining two galaxies 
were observed with X-Shooter, but no H$\alpha$ was detected indicating that these are at $z>0.1$ and therefore likely background AGN. As such, we find no sources with similar observed optical and radio properties as the host galaxy of FRB~121102 out to $z=0.06$. 

\section{Conclusion}

We have searched for a potential host galaxy of FRB~171020 and found \galname\ to be the most likely candidate given its low redshift and position close to the center of the error ellipse. 
UV imaging from GALEX and follow-up spectroscopic observations reveal that \galname\ is a low-metallicity galaxy with a star-formation rate of 0.13 M$_\odot$\,yr$^{-1}$, similar to the host galaxy of FRB~121102. However, no compact, persistent radio continuum source is detected above $L_{\rm 2.1 GHz}=3.6\times 10^{19}$ W\,Hz$^{-1}$, $600$ times fainter than the persistent source associated with FRB~121102.
There is no galaxy within the localization uncertainty region that has similar properties to the host galaxy of FRB~121102 at redshifts $z\lesssim0.06$. This suggests that not all Fast Radio Bursts have an associated `persistent' radio source. As such, identifying host galaxies based on the presence of a compact, luminous radio continuum source may not necessarily help in identifying host galaxies of FRBs.\\


\acknowledgments

The authors wish to thank S. Chatterjee, C. Bassa and the anonymous referee for useful discussions and suggestions that helped to improve the paper. This research is based on observations collected at the European Southern Observatory under ESO programme 0101.A-0455(B) (PI: Macquart), as well as on observations obtained as part of program GS-2018A-Q-205 (PI: Tejos) at the Gemini Observatory and C3211 (PI: Shannon) on the Australia Telescope Compact Array. J.P.M., R.M.S. and K.B. acknowledge Australian Research Council grant DP180100857 and R.M.S. acknowledges support through Australian Research Council (ARC) grants FL150100148 and CE17010000.

The Australian SKA Pathfinder is part of the Australia Telescope National Facility which is managed by CSIRO. We acknowledge the Wajarri Yamatji as the traditional owners of the Observatory site. The Australia Telescope Compact Array is part of the Australia Telescope National Facility which is funded by the Australian Government for operation as a National Facility managed by CSIRO. The Gemini Observatory is operated by the Association of Universities for Research in Astronomy, Inc., under a cooperative agreement with the NSF on behalf of the Gemini partnership: the National Science Foundation (United States), the National Research Council (Canada), CONICYT (Chile), Ministerio de Ciencia, Tecnolog\'{i}a e Innovaci\'{o}n Productiva (Argentina), and Minist\'{e}rio da Ci\^{e}ncia, Tecnologia e Inova\c{c}\~{a}o (Brazil). 

%

\facilities{ASKAP, ATCA, WISE, VISTA, GALEX, Gemini-S, VLT}


\software{astropy \citep{astropy}, PyRAF 
          }

\bibliography{frb171020}





\end{document}